\documentclass[final,5p,times,twocolumn,authoryear]{elsarticle}

\usepackage{amssymb}
\usepackage{lipsum}

\usepackage{hyperref}
\hypersetup{colorlinks,citecolor=blue,filecolor=blue,linkcolor=blue,urlcolor=blue}
\usepackage{graphicx}
\usepackage{dcolumn}

\usepackage{multirow}
\usepackage{threeparttable}
\usepackage[T1]{fontenc}

\usepackage{float}

\journal{Physics Letters B}

\begin{document}

\begin{frontmatter}



\title{Quenching of single-particle strength inferred from nucleon-removal transfer reactions on $^{15}$C}

\author[sustech,CIAE]{Y.~C.~Jiang}
\author[sustech,anl]{J.~Chen\corref{cor1}}
\ead{chenjie@sustech.edu.cn}
\author[anl]{B.~P.~Kay}
\author[anl]{C.~R.~Hoffman}
\author[anl]{T.~L.~Tang}
\author[anl]{I.~A.~Tolstukhin}
\author[IMP,UCAS]{M.~R.~Xie}
\author[IMP,UCAS]{J.~G.~Li}
\author[IMP,UCAS]{N.~Michel}
\author[anl]{M.~L.~Avila}
\author[Spain]{Y.~Ayyad}
\author[msu]{D.~Bazin}
\author[Manchester]{S.~Bennett}
\author[anl]{J.~A.~Clark}
\author[Manchester,cern]{S.~J.~Freeman}
\author[anl]{H.~Jayatissa}
\author[PKU]{G.~Li}
\author[sustech,CIAE]{W.~P.~Liu}
\author[PKU]{J.~L.~Lou}
\author[Spain]{A.~Munoz-Ramos}
\author[anl]{C.~Müller-Gatermann}
\author[msu]{T.~Nathan}
\author[anl]{D.~Santiago-Gonzalez}
\author[Manchester]{D.~K.~Sharp}
\author[CIAE]{Y.~P.~Shen}
\author[Connecticut]{A.~H.~Wuosmaa}
\author[Zhuhai]{C.~X.~Yuan}

\affiliation[sustech]{organization={Department of Physics, Southern University of Science and Technology},
           city={Shenzhen},
           postcode={518055}, 
           state={Guangdong},
           country={China}}
\affiliation[CIAE]{organization={China Institute of Atomic Energy},
           city={Beijing},
           postcode={102413},
           country={China}}
\affiliation[anl]{organization={Physics Division, Argonne National Laboratory},
           city={Lemont},
           postcode={60439}, 
           state={Illinois},
           country={USA}}
\affiliation[IMP]{organization={Institute of Modern Physics, Chinese Academy of Sciences},
           city={Lanzhou},
           postcode={730000}, 
           state={Gansu},
           country={China}}
\affiliation[UCAS]{organization={School of Nuclear Science and Technology, University of Chinese Academy of Sciences},
           city={Beijing},
           postcode={100049}, 
           country={China}}
\affiliation[Spain]{organization={IGFAE, Universidade de Santiago de Compostela},
           city={Santiago de Compostela},
           postcode={E-15782}, 
           country={Spain}}
\affiliation[msu]{organization={Facility for Rare Isotope Beams, Michigan State University},
           city={East Lansing},
           postcode={48824}, 
           state={Michigan},
           country={USA}}
\affiliation[Manchester]{organization={Department of Physics, University of Manchester},
           city={Manchester},
           postcode={M13 9PL}, 
           country={United Kingdom}}
\affiliation[cern]{organization={EP Department, CERN},
           city={Geneva},
           postcode={CH-1211}, 
           country={Switzerland}}
\affiliation[PKU]{organization={School of Physics and State Key Laboratory of Nuclear Physics and Technology, Peking University},
           city={Beijing},
           postcode={100871},
           country={China}}
\affiliation[Connecticut]{organization={Department of Physics, University of Connecticut},
           city={Storrs},
           postcode={06269}, 
           state={Connecticut},
           country={USA}}
\affiliation[Zhuhai]{organization={Sino-French Institute of Nuclear Engineering and Technology, Sun Yat-Sen University},
           city={Zhuhai},
           postcode={519082}, 
           state={Guangdong},
           country={China}}

\cortext[cor1]{Corresponding Author}

\begin{abstract}
The difference in the proton and neutron separation energies ($\Delta S$) of the weakly bound $^{15}$C ground state is -19.86 MeV, an extreme value. Data from intermediate-energy heavy-ion induced (HI-induced) knockout reactions on nuclei spanning $-20\lesssim\Delta S\lesssim+20$ MeV, suggest that the degree to which single-particle strength is quenched, $R\mathrm{_{s}}$, has a negative correlation with $\Delta S$, decreasing from unity around $-20$~MeV to around 0.2 at $+20$~MeV. For the $^{15}$C ground state ($R_s=0.96(4)$ in HI-induced knockout), contrasting results have recently been obtained via the neutron-adding transfer reaction, which reveal a value of $R_s=0.64(15)$, similar to the value observed at modest $\Delta S$ and more extreme values of $\Delta S$ with reaction probes other than HI knockout. In order to explore the any potential differences between $adding$ and $removing$ processes in transfer reactions at extreme $\Delta S$, single-neutron removal transfer reactions on $^{15}$C were performed at 7.1MeV/u in inverse kinematics. The removal of a valence neutron in 2$s_{1/2}$ orbit using both ($p$,$d$) and ($d$,$t$) reactions shows consistent quenching factors and agrees with those from the neutron-adding reaction. The present results, which can be compared with neutron knockout reaction, suggest that correlations, represented by the quenching factor, show limited dependence on neutron-proton asymmetry under the most extreme asymmetry conditions so far achieved in transfer reactions.

\end{abstract}




\end{frontmatter}




\section{Introduction}
\label{introduction}

The independent single-particle model as a tool to interpret quantum many-body systems of nucleons has proven surprisingly successful~\cite{Pandharipande_RMP_1997, Hen_RMP_2017}. However, a more realistic description of nuclei requires the correlations among nucleons occurring at short~\cite{Piasetzky_PRL_2006,SUBEDI_Science_2008} and long~\cite{Dickhoff_PPNP_2004,Barbieri_PRL_2009} distances, which have importance implications on the nature of dense nuclear matter and in understanding systems such as neutron stars \cite{SUBEDI_Science_2008, Duer_Nature_2018}. Data from lepton-induced knockout reactions, ($e$,$e'p$), provided clear evidence for such correlations, revealing a ``quenching'' of the single-particle strength of stable nuclei from Li to Pb \cite{Lapikás_NPA_1993,Kramer_NPA_2001}. The single-particle strengths, quantified by the reaction cross sections, were found to be reduced to around 60\% compared to independent single-particle model predictions and appear to be independent of mass and orbital angular momentum. This reduction factor of the single-particle strengths is the so-called quenching factor $R\mathrm{_{s}}$, which can simply be defined as the ratio of the experimental cross section to the theoretical one (for which the theoretical spectroscopic factor (SF) is considered)~\footnote{Since the cross sections are directly related to the SFs, $R\mathrm{_{s}}$ can also be derived from calculating the ratio of the experimental SF (SF$\mathrm{_{exp}}$) to the SF calculated by shell-model (SF$\mathrm{_{SM}}$).}. It acts as an indicator of the degree to which nucleons are  correlated in the nuclear medium.

\begin{figure*}[htb]
    \centerline{\includegraphics[width=1.5\columnwidth]{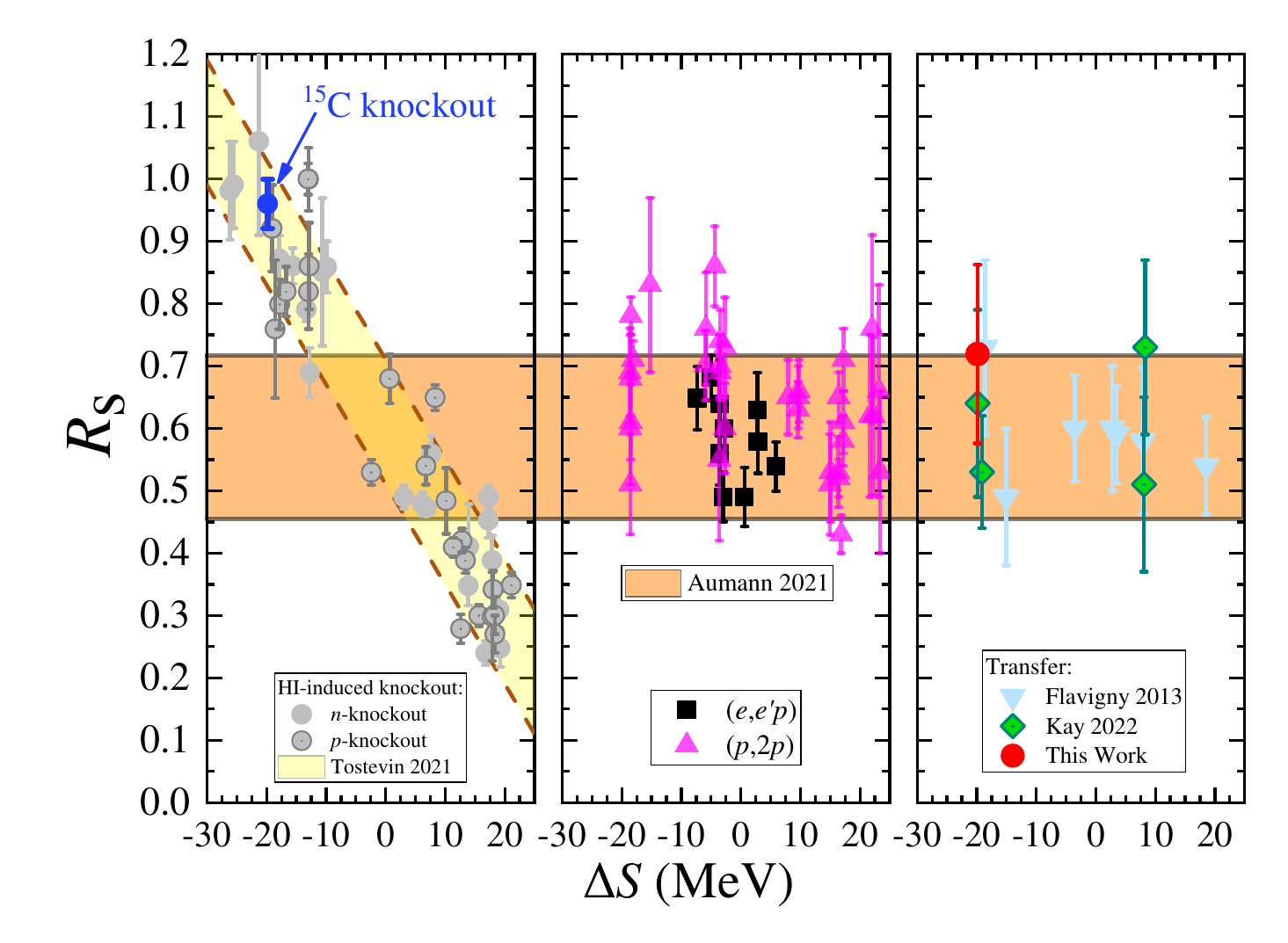}}
    \caption{\label{fig1} 
    Quenching factor $R_{s}$ as a function of the neutron-proton asymmetry $\Delta S$ (for definitions, see text). The orange horizontal band is the same as that of Fig. 56 of Ref. \cite{Aumann_PPNP_2021} and is to guide the eye. The data points in the first and second panel are derived from HI-induced knockout \cite{Tostevin_PRC_2021} (grey circles), ($e$,$e'p$) \cite{Kramer_NPA_2001,Tostevin_PRC_2014} (black squares), and proton-induced ($p$,$2p$) knockout \cite{Atar_PRL_2018,Kawase_PTEP_2018,Ramos_PLB_2018,Holl_PLB_2019,Pohl_PRL_2023,Wamers_PRC_2024} (magenta triangles). Note that the HI-induced $^{15}$C neutron knockout data 0.96(4) (first published as 0.90(4)~\cite{Terry_PRC_2004} but subsequently revised to be 0.96(4)~\cite{Lee_PRC_2006} in works that followed.) is specially marked. The yellow band in panel 1 shows the trend of HI-induced knockout from the compilation of Ref. \cite{Tostevin_PRC_2021}. The light-blue points in the third panel show the quenching factors of oxygen isotopes \cite{Flavigny_PRL_2013}. The green diamonds show recent $^{15}$C and $^{15}$N data \cite{Kay_PRL_2022}. The results of this work (red circles) are consistent with the ground state $^{15}$C quenching factor from transfer \cite{Kay_PRL_2022}, and differs from $^{9}$Be-induced $^{15}$C knockout \cite{Terry_PRC_2004}.
    }
\end{figure*}

Heavy ion induced (HI-induced) one-nucleon knockout reactions from intermediate-energy projectile beams have proved to be a useful probe to extract single-particle strength, particularly in exotic systems \cite{Hansen_ARNPS_2003}. It is generally assumed that the short-range properties of nuclei should not depend on reaction mechanisms. However, systematic studies of HI-induced knockout experiments have drawn a surprising conclusion, that the quenching factor $R_{s}$ strongly depends on the Fermi surface asymmetry \rm{$\Delta$}$S$ \cite{Tostevin_PRC_2014,Tostevin_PRC_2021,Aumann_PPNP_2021}, where \rm{$\Delta$}$S$ is defined as ($S$\rm{$_{p}$}-$S$\rm{$_{n}$})/($S$\rm{$_{n}$}-$S$\rm{$_{p}$}) for proton/neutron removal to the g.s., respectively (see Fig. \ref{fig1} for details). Despite some recent HI-induced knockout experiments suggesting a slightly dampened slope, the asymmetry dependence remains a puzzle~\cite{Sun_PRC_2021,Sun_PRC_2022, Sun_PRC_2024}. In contrast, recent $proton$-induced quasi-free knockout ($p$,$2p$) measurements have shown a very limited asymmetry dependence for $R_{s}$ \cite{Atar_PRL_2018,Kawase_PTEP_2018,Ramos_PLB_2018,Holl_PLB_2019,Pohl_PRL_2023,Wamers_PRC_2024}, making this long-standing issue even more complicated and motivating a need to better understand the different reaction mechanism, experimentally and theoretically.

Transfer reactions provide another approach to study single-particle strength. No asymmetry dependence has been found from transfer reactions on stable nuclei so far, consistent with the $electron$- or $proton$- induced knockout data \cite{Kramer_NPA_2001}. In recent years, the study of the quenching effect using inverse-kinematic transfer reactions at radioactive ion beam facilities supported the systematic trends of a weak dependence of the $R_{s}$ on $\Delta S$ deduced from previous studies on stable nuclei. For example, experiments using the $^{34,36,46}$Ar isotopes (covering a range of -10 $\leq$ \rm{$\Delta$}$S$ $\leq$ 12 MeV) \cite{Lee_PRL_2010,Manfredi_PRC_2021} have also shown much weaker quenching factor asymmetry dependence compared to HI-induced knockout. However, quantitative experimental data from transfer reactions on extremely deeply bound or weakly bound nuclei are lacking, especially for those with \rm{$\Delta$}$S$ $\leq$ -15 MeV, where the discrepancies are most significant (See Fig. \ref{fig1} for details). The only two recent experiments that have explored the quenching puzzle in such exotic systems are the studies of the $^{14}$O \cite{Flavigny_PRL_2013} and $^{15}$C nuclei \cite{Kay_PRL_2022}.

The nucleus $^{15}$C can be viewed as an inert $^{14}$C core plus a valence neutron in the 2$s_{1/2}$ orbit. The spectroscopic factor calculated\footnote{The correction made to account for the isobaric analog state (IAS) is negligible, i.e., the isospin-coupling Clebsch-Gordan coefficient \textit{C}$^{2}$ is 1 in our case.} using the shell model is very close to unity, the independent single-particle limit. Examining the quenching effect in nuclear systems with extreme $\Delta S$ such as $^{15}$C can provide valuable insights to aid our understanding of this long-standing issue. A recent measurement of the $^{14}$C($d$,$p$)$^{15}$C \cite{Kay_PRL_2022} shows a discrepancy of approximately 35\% in the quenching factor compared to the HI-induced knockout of $^{15}$C \cite{Terry_PRC_2004}. It is not clear, however, whether such a substantial discrepancy arises from differences in the intrinsic reaction mechanisms between knockout and transfer, or whether there are intrinsic differences between nucleon $adding$ and $removing$ processes. Therefore, more experiments using different probes are needed to verify the current discrepancies, particularly those focusing on the nucleon $removal$ reactions, where the reaction direction is the same as in HI-induced knockout.

To clarify the potential differences between $adding$ and $removing$ processes in transfer reactions related to $^{15}$C, we used the weakly bound $^{15}$C nucleus to study single-neutron removal transfer reactions in inverse kinematics. The goal of our work is to verify whether consistent quenching factors in systems that arrive at the same final state, and also hold a reaction direction consistent with HI-induced knockout, are revealed.

In this Letter, we report a new set of experimental data, also shown in Fig. \ref{fig1}, from single-neutron removal transfer reactions $^{15}$C($p$,$d$)$^{14}$C and $^{15}$C($d$,$t$)$^{14}$C. We determined quenching factors $R\mathrm{_{s}}$ consistent with the previous $^{14}$C($d$,$p$)$^{15}$C result \cite{Kay_PRL_2022}. Our new result, derived from a nucleon $removal$ reaction direction consistent with the knockout reaction, emphasizes the current discrepancy between transfer and knockout in the most extreme $\Delta S$ region (-19.86 MeV) achieved so far in transfer reactions. This reaffirms the observation that the strong dependence of $R\mathrm{_{s}}$ on $\Delta S$ is present only in HI-induced knockout reactions.

\section{Experiment}
\label{Experiment}

The experiment was carried out using the HELIOS spectrometer \cite{Lighthall_NimA_2010} at the ATLAS facility at the Argonne National Laboratory. A 7.1 MeV/u $^{15}$C secondary beam was produced at the ATLAS in-flight system, RAISOR \cite{Hoffman_NimA_2022} via the neutron-adding ($d$,$p$) reaction on a $^{14}$C primary beam at 8 MeV/u, with an intensity of 200 pnA. The resulting $^{15}$C beam had a rate of approximately 10$^{6}$ particles per second with negligible contamination (less than 1\%). The secondary beam bombarded a target of either deuterated polyethylene (CD$_{2}$)$_{n}$ or polyethylene (CH$_{2}$)$_{n}$ of thickness 363(20) $\mu$g/cm$^{2}$ and 387(22) $\mu$g/cm$^{2}$ respectively. The target thickness uncertainty is negligible compared to the uncertainties caused by optical potentials and other considerations in the analysis as discussed below.

\begin{figure*}[htb]
    \center{\includegraphics[width=1.65\columnwidth] {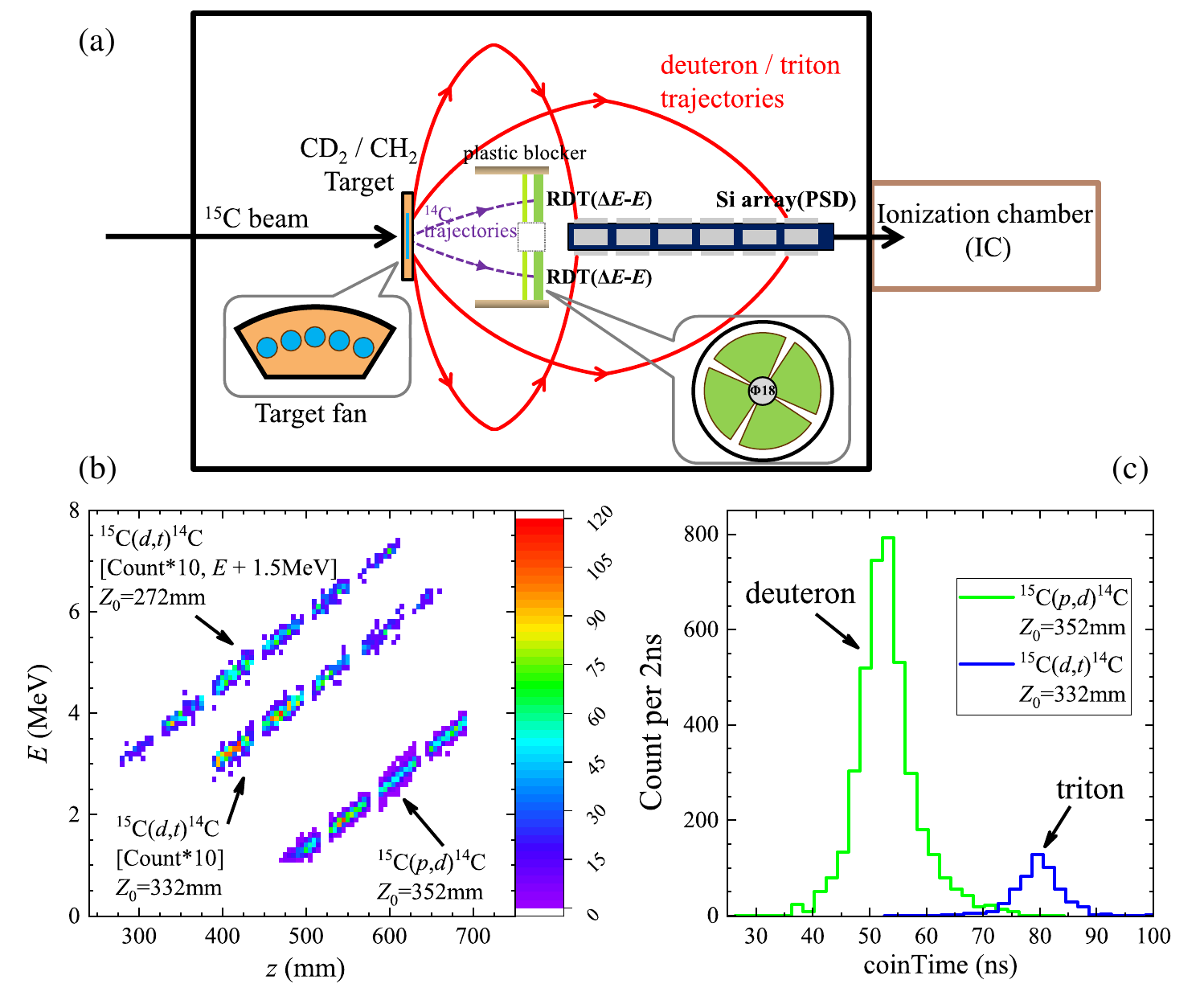}}
    \caption{\label{fig2} 
    (a) A schematic of the experimental setup for $^{15}$C experiment.
    (b) Characteristic kinematics lines for ($p$,$d$) and ($d$,$t$) neutron removing reactions (only the kinematic lines corresponding to the g.s. of $^{14}$C are shown, see text). Note that the kinematic line from the $z$ = 272 mm setting is shifted in energy to avoid overlap. Also, the counts from both settings in ($d$,$t$) measurement have been multiplied by a factor of 10 for better visualization.
    (c) Coincidence Time (coinTime) for different reactions. See text for details.
    }
\end{figure*}

The outgoing deuterons and tritons were detected using the HELIOS solenoid spectrometer in a magnetic field with a strength of 2.5 T. A complete schematic of the experimental setup is shown in Fig. \ref{fig2} (a). The silicon array, composed of 24 position-sensitive detectors (PSD), was placed downstream of the target (where $z=0$), covering a range of 332 mm $\leq$ $z$ $\leq$ 682 mm for the measurement of ($d$,$t$) reaction, and 352 mm $\leq$ $z$ $\leq$ 702 mm for ($p$,$d$) reaction. For the ($d$,$t$) reaction measurement, the target was moved 60 mm closer to the silicon array for part of the measurement to cover smaller center of mass angles. The differential cross sections of the ($d$,$t$) reaction deduced from two settings were cross-checked with each other. The heavy ion residues were measured by the recoil detector telescopes (RDTs) $\sim$ 62 mm upstream of the silicon array. The RDTs consist of four sets of $\Delta E$-$E$ telescopes composed of quadrant silicon detectors with thicknesses of $\approx$ 75 $\mu$m and 1000 $\mu$m, respectively. Four RDTs were arranged leaving a 18 mm diameter hole in the center for the beam to pass through. In addition, a cylindrical plastic blocker surrounding the RDTs was designed to block the protons, deuterons and tritons traveling for more than one cyclotron period. The time difference between silicon detector and RDT was used to analyze the coincidence time and for particle identification.

The incident beam was monitored by a fast-counting ionization chamber (IC) \cite{Lai_NimA_2018} located $\approx$ 1000 mm downstream of the target. The rate in the IC calculated by a discriminator threshold was used as a scaler to count the total incident ions. Also, the beam intensity was verified based on ($p$,$p$) or ($d$,$d$) elastic scattering data \cite{Chen_PRC_2022}. An uncertainty of 15\% for beam intensity was taken into account for the quenching analysis below.

\section{Data Analysis and Results}
\label{Data Analysis and Results}

Since the cyclotron period of charged particles in a magnetic field is proportional to the mass-charge ratio, the time difference between silicon array and RDT was used for particle identification. The particle-identification gate on the \rm{$\Delta$}$E$-$E$ spectra and coincidence time was verified using a kinematics simulation \cite{Tang_zenodo_2019}. With these gates, the characteristic kinematic lines for both $(d,t)$ and $(p,d)$ reaction channels are plotted in Fig. \ref{fig2} (b). The energy of the outgoing light particles $E_{\mathrm{lab}}$ exhibits a linear relationship with the $z$ position, which was deduced from the difference in signal amplitudes from the two ends of the PSD \cite{Hoffman_PRC_2012}. It should be noted that only the reactions to the g.s. of $^{14}$C should give kinematic lines in Fig. \ref{fig2} (b) because of the acceptance of the spectrometer, and the higher-lying excited states are outside of this acceptance.

The spectrum of coincidence time between the silicon array and the RDT corresponding to Fig. \ref{fig2} (b) is shown in Fig. \ref{fig2} (c). For simplicity, only one setting in the ($d$,$t$) measurement is displayed. The cyclotron periods for deuterons and tritons under a magnetic field with a strength of 2.5 T are approximately 52.6 and 78.9 ns, respectively. The peak positions shown in Fig. \ref{fig2} (c) match these periods as expected, which further confirms the particle identification.

\begin{figure*}[htb]
    \center{\includegraphics[width=1.6\columnwidth]{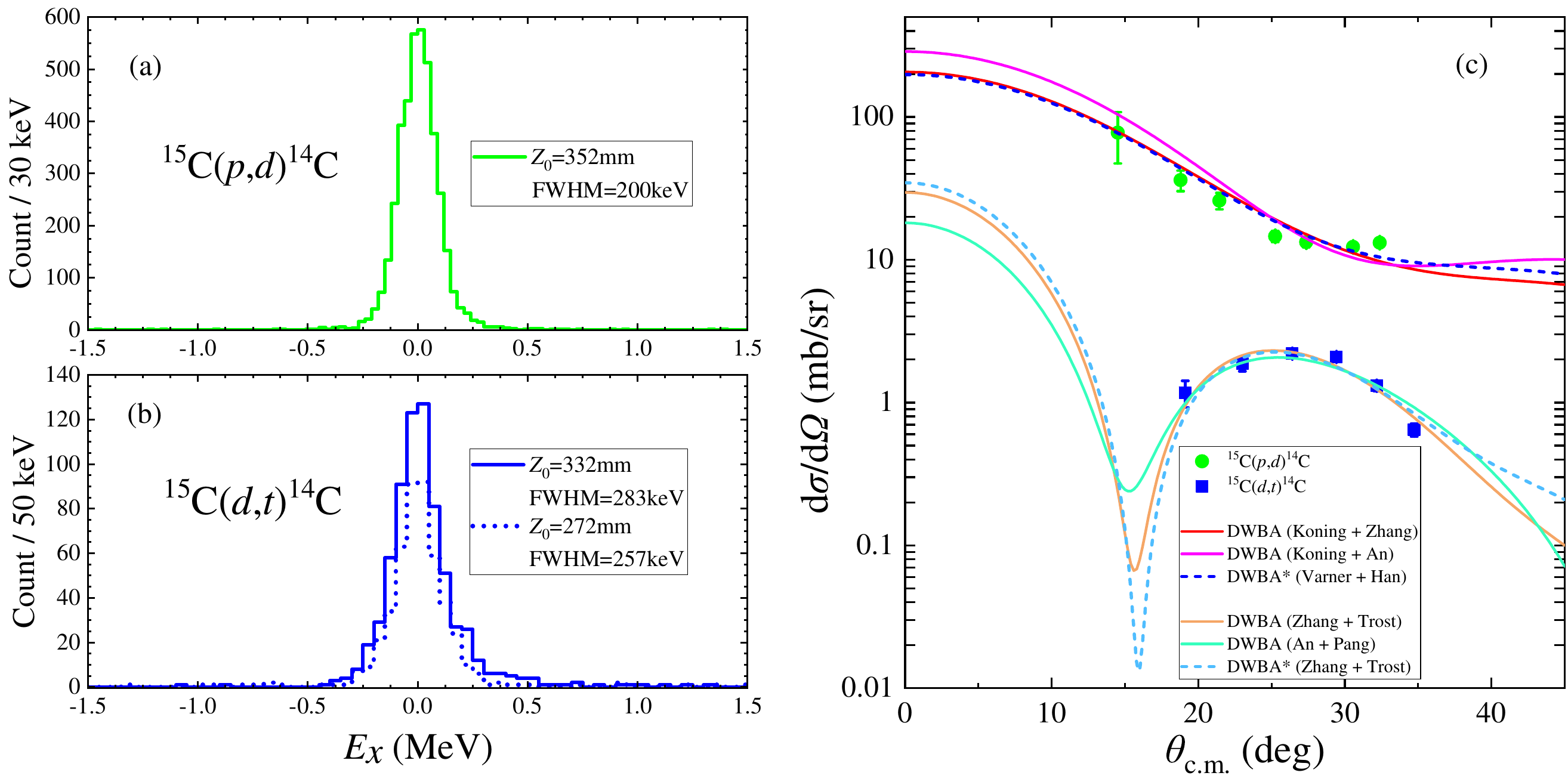}}
    \caption{\label{fig3} 
    (a)\&(b) The excitation spectra for the ($p$,$d$) and ($d$,$t$) reactions respectively, where the FWHM of the peaks show the experimental resolution.
    (c) Differential cross sections for $^{15}$C($p$,$d$)$^{14}$C (green circles) and $^{15}$C($d$,$t$)$^{14}$C (blue squares). DWBA calculations have been shown as well, where the solid lines represent those calculated using global optical potentials and the dashed lines represent those calculated using local potentials.}
\end{figure*}

With the particles of interest identified, the spectrum of excitation energy was reconstructed according to the reaction kinematics, which confirmed that the g.s. of $^{14}$C was populated via one-neutron removal of $^{15}$C (see Fig. \ref{fig3} (a) \& (b)). The width of the ground state peak illustrates the excitation-energy resolution of 200-300 keV FWHM. The center-of-mass angle $\theta_{\mathrm{c.m.}}$ was determined using the $z$ position. The data from each individual PSD were divided into two or more bins as statistics allowed. The final differential cross-sections are shown in Fig. \ref{fig3} (c). For ($p$,$d$) reaction, the uncertainty of the first data point is relatively large because of the acceptance correction resulting from the 18-mm diameter hole in the center of the RDTs (see Fig. \ref{fig2} (a)) , but had a minor impact on the spectroscopic factor determination, as discussed below.

The 18-mm hole in the center of the RDTs led to some loss of recoil detection. Taking into account various beam conditions and calculating the proportion of events with signal on the silicon array but no signal on the RDT in the simulation, the corrected acceptance was determined. For smaller center-of-mass angles ($\leq$ 20$^{\circ}$), the corrections are relatively large, especially for ($p$,$d$) reaction. For larger center-of-mass angles, the correction was within 14\%. Also, as the gate of coincidence time did not fully capture all events, about 10\% of the events fell outside the timing gate and were not captured by the aforementioned particle identification. This portion has also been included in the angular distribution correction. A 50\% uncertainty has been adopted to both corrections as a conservative estimation (e.g., a 5\% uncertainty was added to the 10\% correction). The error bars shown in Fig. \ref{fig3} (c) include all the above mentioned uncertainties.

To extract spectroscopic factors from the experimental data, the Distorted-Wave Born Approximation (DWBA) method was used, which has been well validated over many decades (for example, see Ref.~\cite{Schiffer_PRL_2012, Schiffer_PRC_2013, Kay_PRL_2013}). The DWBA calculations were done using the finite-range code PTOLEMY\footnote{Note that the ``cancellation hypothesis'' in Ref.~\cite{Timofeyuk_PRC_1999} is still valid even in the halo nucleus $^{15}$C and the uncertainty of SF resulting from this hypothesis is negligible compared to that induced by optical potentials.} \cite{Macfarlane_1978}. The $^{15}$C bound-state form factors were generated using a Woods-Saxon (WS) potential, defined by $r\mathrm{_{0}}$ = 1.25 fm, $a\mathrm{_{0}}$ = 0.65 fm, $V\mathrm{_{so}}$ = 6 MeV, $r\mathrm{_{so0}}$ = 1.1 fm, and $a\mathrm{_{so0}}$ = 0.65 fm. The depth of the potential was varied to reproduce the binding energy of the transferred nucleon to the final state. The deuteron and triton bound states were taken from Ref. 
\cite{Wiringa_PRC_1995} and Ref. \cite{Brida_PRC_2011}, respectively.

Different global optical-model parameterizations were explored in the incoming and outgoing channels for both reactions. Given that most of the existing global optical potentials for hydrogen isotopes are derived from scattering experiments on isotopes with larger $A$ values and relatively high energies (typically $\geq$ 10 MeV), we chose global optical potentials which cover mass range close to $A$=15 and cover lower energies. For protons and deuterons, the parameterizations of Ref. \cite{Koning_NPA_2003,Varner_PR_1991,Perey_PR_1976} and Ref. \cite{An_PRC_2006,Han_PRC_2006,Zhang_PRC_2016,Daehnick_PRC_1980} were used, respectively. Our choice is similar to that of Ref. \cite{Kay_PRL_2022}. Also, we note that Cole {\it{et al}}. have carried out an elastic-scattering measurement for $^{14}$C($d$,$d$)$^{14}$C close to the present beam energy~\cite{Cole_NPA_1973}, so this set of optical potential parameters was also included for the exit channel of the $^{15}$C$(p,d)^{14}$C reaction. The spectroscopic factor derived using the potential from Cole {\it{et al}}. differs by less than 3\% from the average of all spectroscopic factors determined using other global optical potentials, further validating our optical potential choice. For tritons, due to the scarcity of data on triton scattering, we considered using global optical potentials derived from scattering information of $A$=3 nuclei. As a result, those from Ref. \cite{Xu_SCPMA_2011,Pang_PRC_2009,Liang_JPG_2009,Trost_NPA_1987} were taken into account.

The experimental spectroscopic factors were derived from fitting the experimental angular distributions to the DWBA calculations, as shown in Fig.~\ref{fig3} (c) by the solid/dashed lines. The quenching factor ($R\mathrm{_{s}}$) is derived by calculating the ratio of SF$\mathrm{_{exp}}$ to SF$\mathrm{_{SM}}$, i.e., SF$\mathrm{_{exp}}$/SF$\mathrm{_{SM}}$. The later was calculated with the YSOX interaction within the $psd$-shell \cite{Yuan_PRC_2012}. We found that the standard deviation of the spectroscopic factor or quenching factor is approximately 14\% for ($p$,$d$) and 24\% for ($d$,$t$), where the larger uncertainty is caused by the greater ambiguity of optical potentials in the triton-related reaction.

In addition to calculating $R\mathrm{_{s}}$ using global optical potentials, we also fitted these optical potentials to the elastic scattering data using the potential normalization method described in \cite{Chen_PRC_2022,Chen_PRC_2016a,Chen_PRC_2016b} with the code SFRESCO \cite{Thompson_CPR_1988}, to further constrain the potentials. With these local potentials, new $R\mathrm{_{s}}$ values were obtained (see also the dashed lines in Fig. \ref{fig3} (c)). The spectroscopic factors with the global and local potentials show an acceptable degree of difference, 18\% for ($p$,$d$) reaction and 9\% for ($d$,$t$) reaction, respectively. The spectroscopic factors derived from global and local potentials have been averaged for the final results shown in Table \ref{tab:Rs}. Note that the uncertainties include the standard deviation of SF fitting using local and global optical potential parameter sets, uncertainties of the target thickness and beam intensity. 

Indeed, it has been established for a long time that the selection of bound-state potential parameters is crucial for extracting spectroscopic factors, whether in knockout reactions (see, e.g., Ref.~\cite{Brown_PRC_2002}), transfer reactions (see, e.g., Ref.~\cite{Lee_PRC_2006,Mukhamedzhanov_PRC_2005}) or Coulomb dissociation reactions (see, e.g., Ref.~\cite{Nakamura_PRC_2009}). For example, spectroscopic factors more consistent with those deduced from ($e$,$e'p$) reactions can be obtained from transfer reactions if the radius parameters $r_{0}$ are constrained by Hartree-Fock calculations~\cite{Hai_PRC_2024}. Therefore, we also tested the $r_0$ sensitivity by using the recommended single-particle potential parameter of $r\mathrm{_0}$ = 1.34~\cite{Hai_PRC_2024}. The results were approximately 3\% smaller than those obtained with $r\mathrm{_0}$ = 1.25, which may be related to the weakly bound nature of the $s_{1/2}$ orbit, but require further investigations.
All the $R\mathrm{_{s}}$ determined under different groups of potentials are given in Table \ref{tab:Rs}. The uncertainty resulting from the variation of bound-state parameter $a_0$ is also much smaller than the one resulting from the different optical model parameters.


Table \ref{tab:Rs} shows that the results from the two reaction channels are self-consistent and agree with the ground state results for $^{15}$C reported by Kay {\it et al}. \cite{Kay_PRL_2022}. $R_{s}$= 0.72$\pm$0.14 was deduced by averaging the results of two reaction channels (see the first two rows in Table \ref{tab:Rs}) and plotted in Fig.~\ref{fig1} (red circle), since they are probing the same spectroscopic factor. The present result contradicts the quenching factor (0.96$\pm$0.04) deduced from the HI-induced knockout reaction~\cite{Terry_PRC_2004} (see Fig.~\ref{fig1}).

\begin{table}
    \centering
    \caption{Spectroscopic factors derived from different parameter sets, and corresponding deduced $R\mathrm{_{s}}$ for valence neutron strength of $^{15}$C ground state.\label{tab:Rs}}
    \resizebox{0.95\columnwidth}{!}{%
        \begin{tabular}{ccccccc}
        \hline
        $^{A}X$ & $nlj$ & Reaction & Bound State $r\mathrm{_0}$ & SF$\mathrm{_{exp}}$ & SF$\mathrm{_{SM}}$ & $R\mathrm{_{s}}$ \\
        \hline
        \multirow{4}{*}{$^{15}$C} & \multirow{4}{*}{2$s_{1/2}$} & ($p$,$d$) & $r\mathrm{_0}$ = 1.25 & 0.67(18) & 0.95 & 0.70(19) \\
        &  & ($d$,$t$) & $r\mathrm{_0}$ = 1.25 & 0.70(20) & 0.95 & 0.74(21) \\
        &  & ($p$,$d$) & $r\mathrm{_0}$ = 1.34 & 0.65(18) & 0.95 & 0.68(19) \\
        &  & ($d$,$t$) & $r\mathrm{_0}$ = 1.34 & 0.68(20) & 0.95 & 0.72(21) \\
        \hline
        \end{tabular}
    }
\end{table}

\section{Discussion}
\label{Discussion}

As mentioned above, $^{15}$C is a nucleus with both large negative $\Delta S$ and well-defined single-neutron configuration with an inert $^{14}$C core. From the independent-particle model, the spectroscopic factor should be unity, which is close to the result from the shell-model calculation. The present experimental spectroscopic factor ($0.68\pm0.14$) is much smaller than both, clearly revealing the quenching effect in the extremely weakly bound system $^{15}$C. 

One may question the difference between the inclusive analysis for all bound states under knockout reactions and the exclusive analysis for a single state under transfer reactions. Studies of transfer reactions on stable nuclei seem to confirm that the quenching factor derived from transfer reactions is independent of the orbital angular momentum $\ell$ of the final state and consistent over a wide mass range~\cite{Aumann_PPNP_2021, Kay_PRL_2013}. Furthermore, Xu {\it{et al}}. have recently re-analyzed a large number of ($p$,$d$) transfer data for both stable and unstable nuclei \cite{Xu_PLB_2019}, suggesting that inclusive and exclusive treatments are not the reason of the ``quenching puzzle".

It should also be noted that the reported spectroscopic factor of HI-induced knockout reaction by Terry {\it {et al}}. was deduced from the knockout of a neutron in the $^{15}$C ground state leading to the $^{14}$C ground state. This is the similar reaction direction as the present $removal$ reaction, and thus provides an important insight for clarifying current disagreement between transfer and knockout.

To investigate if the quenching effect can be incorporated by the different theoretical models, we have obtained theoretical spectroscopic factors with the Gamow shell model (GSM) \cite{Xie_PLB_2023, Xie_SCPMA_2024} and $ab$ $initio$ no-core shell model (NCSM) \cite{Wang_PRC_2024}. The GSM enables many-body computations using the Gamow-Berggren basis in rigged Hilbert spaces \cite{Berggren_NPA_1968, Michel_JPG_2009, Li_Physics_2021, Michel_Springer_2021}. In the GSM calculation, a core-plus-valence-nucleon framework was employed, with $^{8}$He chosen as the core and the continuum-coupling effect incorporated. To reduce computational complexity, no more than two particles were allowed to occupy the continuum. The Minnesota phenomenological interaction was employed. The GSM result (0.97) is similar to that of the standard SM and the quenching effect is not well described. This indicates that the quenching effect cannot be captured by just considering the continuum-coupling effect. 

\begin{figure}[htb]
    \center{\includegraphics[width=0.8\columnwidth]{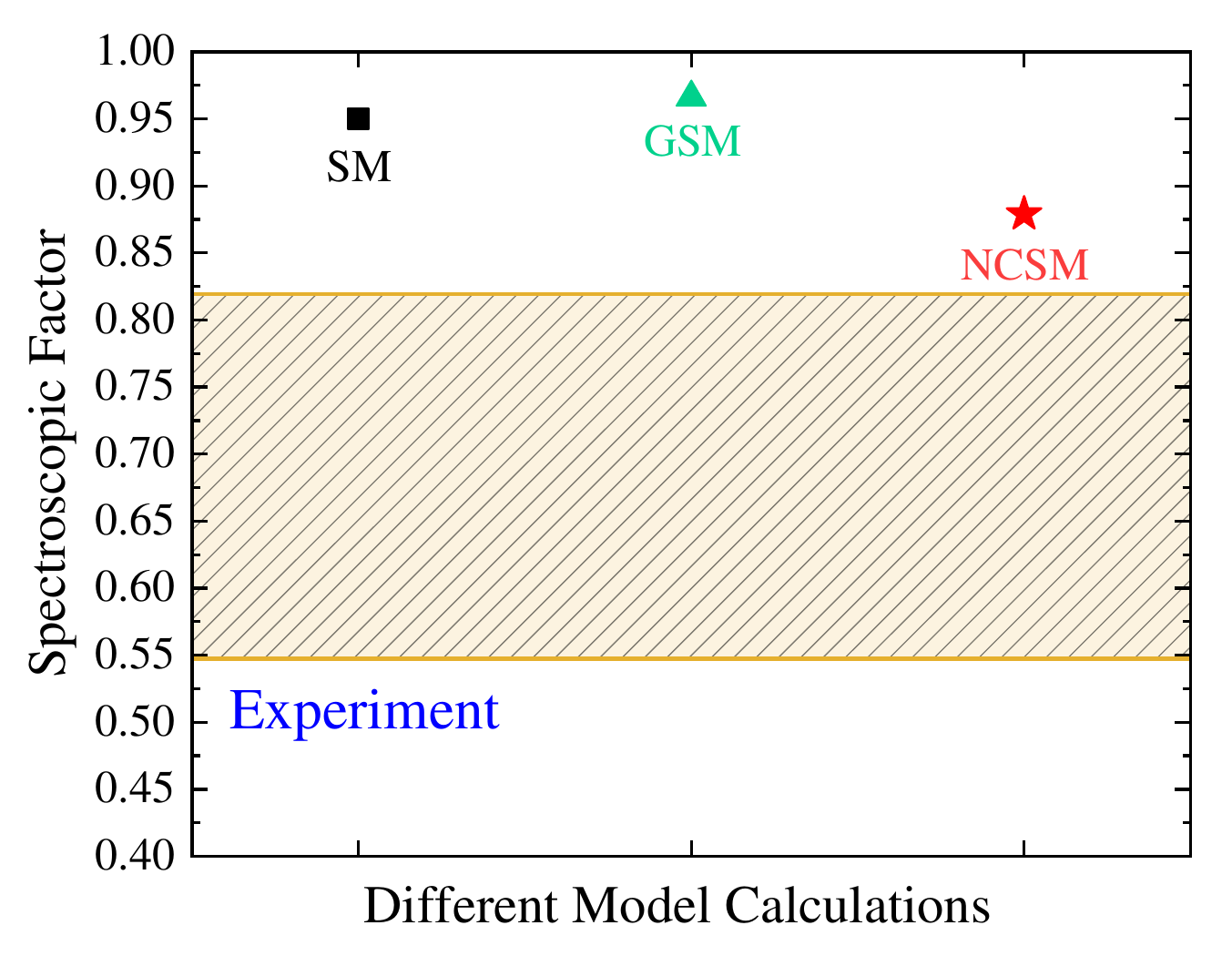}}
    \caption{\label{fig4} 
    Comparison of our experimental SF (shaded band) with theoretical SFs, including SM, GSM and NCSM. It suggests that NCSM agrees better with our experimental value.}
\end{figure}

The NCSM relies on realistic nuclear interactions and enables a self-consistent solution of the nuclear many-body Hamiltonian \cite{Wang_PRC_2024}. We used the Daejeon16 realistic interaction derived from the chiral two-body interaction (N3LO) through the similarity renormalization group (SRG) and unitary transformation methods. This interaction can provide an accurate description of the properties of light nuclei without the need to explicitly include three-body forces \cite{Shirokov_PLB_2016}. In the NCSM, the dimensionality of calculations increases rapidly with the nuclear mass, necessitating the use of progressively larger model spaces, denoted by $N_\mathrm{max}$, which presents the total number of oscillator quanta allowed above the minimum for a given nucleus in the many-body HO basis. Due to computational constraints, calculations were limited to $N_\mathrm{max}$ = 6 with $\hbar \omega$ = 15 MeV. The ground-state spectroscopic factor calculated by NCSM is 0.88, which is about 7\% lower than that of the standard SM. This value agrees better with our experimental spectroscopic factor as shown in Fig. \ref{fig4}. This is probably because NCSM utilizes a more advanced computational approach and realistic interactions that will better describe the correlations within the nucleus. The NCSM framework shows a similar reduction of SFs in some other nuclei \cite{Sun_PRC_2024}. Although current computational limitations prevent us from fully reproducing the observed quenching factor, it is anticipated that the correlations in the $^{15}$C nucleus may be fully captured with the introduction of a more complete model space.

More transfer-reaction data with extreme $\Delta S$ and reduced quantitative uncertainties are required, in particular the $removal$ transfer reactions, to yield a more comprehensive picture of single-particle strength and nucleon-nucleon correlations. It is anticipated that more data will become available as radioactive beams are becoming available for transfer reaction studies, such as at the Beijing Radioactive Ion beam Facility (BRIF) in China, the Isotope Separator On Line DEvice (ISOLDE) in Europe, and the Facility for Rare Isotope Beams (FRIB) in the USA. This makes it feasible to carry out a systematic study of nucleon-$adding$ and $removal$ reactions in the future, which will shed light on future theoretical solutions to the discrepancy of quenching from different probes or reactions.

Efforts from the theoretical end are also crucial. For example, the coupling to the continuum for weakly bound systems like deuteron and $^{15}$C can be taken into account using the continuum discretized coupled channel (CDCC) method (see, e.g., Ref.~\cite{Austern_PR_1987,Gallardo_PRC_2009,Yahiro_PTEP_2012}). Furthermore, there are still significant uncertainties in the optical potentials and different reaction theoretical frameworks. Constraining these uncertainties is another key to solving this problem that has plagued us for nearly 20 years. Noticeable progress has been made recently towards this direction (see, e.g., Ref.~\cite{Lovell_JPG_2021, Pruitt_PRC_2023, Hebborn_PRC_2023, Hebborn_PRL_2023,Hebborn_JPG_2023}).


\section{Summary}
\label{Summary}

In conclusion, the quenching factors of the 2$s_{1/2}$ valence neutron of $^{15}$C have been derived from two neutron $removal$ reactions ($p$,$d$) and ($d$,$t$) consistently, which can be compared with HI-induced neutron knockout. This  complements the extremely limited data on transfer reactions involving weakly bound nuclei and reduces potential uncertainties from differences between $adding$ and $removing$ processes, thus further supporting the hypothesis that the quenching factor $R\mathrm{_{s}}$ is nearly independent of neutron-proton asymmetry $\Delta S$ within the framework of transfer reactions. The new results examine the nature of nucleon-nucleon correlations under the most extreme asymmetry conditions so far achieved for transfer reactions, while posing questions about the reaction mechanisms of transfer and knockout probes.

\section{Acknowledgements}
\label{Acknowledgements}


The authors would like to acknowledge the efforts of the support and operations staff at ATLAS. This research used resources of Argonne National Laboratory’s ATLAS facility, which is a Department of Energy Office of Science User Facility. This material is based upon work supported by the U.S. Department of Energy, Office of Science, Office of Nuclear Physics, under Contracts No. DE-AC02-06CH11357 (ANL), No. DE-SC0020451 (FRIB), No. DE-FG02-87ER40371, No. DE-SC0014552 (UCONN), No. DE-SC0018223 (SciDAC-4/NUCLEI); the Spanish Ministerio de Economía y Competitividad through the Programmes “Ramón y Cajal” with the grant number RYC2019-028438-I; National Natural Science Foundation of China (Grant number No. 12205340, 12147101, 12475120, 12475129, 12435010). The numerical Gamow shell model and no-core shell model calculations in this paper have been done on Hefei advanced computing center. D.K.S. acknowledges U.K. Science and Technology Facilities Council (Grants No. ST/P004423/1 and No. ST/T004797/1). We gratefully acknowledge use of the Bebop cluster in the Laboratory Computing Resource Center at Argonne National Laboratory.



\end{document}